\documentclass{elsart}
\usepackage{graphicx,amsmath,amsfonts, ecltree, epic, amsmath, amssymb, amsfonts}

\begin{document}
\begin{frontmatter}

\title{Kinetic, time irreversible evolution of the unstable $\pi^{\pm}$ - meson}
\author{S.Eh.Shirmovsky}
\ead{shirmov@ifit.phys.dvgu.ru} \address{ Laboratory of Theoretical Nuclear Physics, Far Eastern National University,
Sukhanov Str. 8, Vladivostok, 690950, Russia}
\date{\today}

\begin{abstract}
In Liouville formalism the expression for density matrix,
determining the time evolution of unstable $\pi^{\pm}$ - meson in
the framework of unified formulation of quantum and kinetic dynamics
is defined. The eigenvalues problem is investigated in the framework
of Prigogine's principles of description of nonequilibrium processes
at microscopic level. The problem was solved on the basis of complex
spectral representation. It was shown that the approach leads to
Pauli master equation for the weakly interacting system.
\end{abstract}
\end{frontmatter}

\section{Introduction}
Let me examine the eigenvalues problem for the Hamiltonian
$H=H_{0}+gV$
\begin{equation}\label{h22}
H\mid\psi_{\alpha}>=\tilde{E}_{\alpha}\mid\psi_{\alpha}>,
\end{equation}
where $H_{0}$ - free Hamiltonian, $V$ - interaction part, $g$ - coupling constant. In the conventional case Hamiltonian
$H$ is a Hermitian operator, $\tilde{E}_{\alpha}$ is a perturbed energy of the state - a real number. It is known that
the usual procedure of equation~\eqref{h22} solution on the basis of perturbation method can lead to the appearance of
the small denominators $1/(E_{\alpha}-E_{\alpha'})$, where $E_{\alpha}$, $E_{\alpha'}$ are the energies corresponding
to the unperturbed situation. Obviously, the divergences can arise at $E_{\alpha}=E_{\alpha'}$. The problem of the
small denominators was determined by Poincare as "the basic problem of dynamics"~\cite{poin}. According to Poincare's
classification the systems, for which the situation can be corrected are called the "integrable" systems~\cite{ps,p}.
In the opposite case the systems are defined as the "non-integrable". Poincare proved that in the general case the
dynamic systems are "non-integrable" ones. The basic question now is - what can we do to avoid Poincare's divergences
in case the system is
"non-integrable"? \\
I. Prigogine and co-workers (Brussels - Austin group) noted that in the general case the satisfactory solution of this
problem is impossible on the basis of the conventional formulation of quantum dynamics. The mechanism of asymmetry
processes in time, which made it possible to accomplish a passage from the reversible dynamics to the irreversible time
evolution was developed for the solution of this problem. The authors of the approach deny the conventional opinion
that the irreversibility appears only at the macroscopic level, while the microscopic level must be described by the
laws, reversed in the time. The method of description of the irreversibility at the quantum level proposed by them
leads to the kinetic, time irreversible equations and determines the connection of quantum mechanics with kinetic
dynamics. The approach allows to solve the problems, which could not be solved in the framework of conventional
classical and quantum mechanics, for example, now we can realize the program of Heisenberg - to solve the eigenvalues
problem for the Poincare's "non-integrable" systems. \\
We examine the situation using the simple Friedrichs model (the model is presented closely to the text of
works~\cite{ps},~\cite{ppt} - \cite{pop3}). Despite the fact that the solution of the problem for the Friedrichs model
is known~\cite{frid} it serves as a good example for the demonstration of the essence of situation. The model describes
interaction of two level atom and electromagnetic field. In the Friedrichs model $\mid1>$ corresponds to the atom in
its bare exited level~\cite{opp},~$\mid{k}>$ corresponds to the bare field mode with the atom in its ground state. The
state $\mid1>$ is coupled to the state $\mid{k}>$
\begin{equation}\label{f}
\begin{split}
&H=H_{0}+g V \\
&=E_{1}\mid 1><1\mid+\sum_{k}E_{k}\mid k><k\mid +
g\sum_{k}V_{k}(\mid k><1\mid + \mid 1><k\mid),
\end{split}
\end{equation}
where
\begin{equation}\label{f2}
\mid 1><1\mid + \sum_{k}\mid k><k\mid =1,~<\alpha\mid\alpha '>=\delta_{\alpha\alpha '}
\end{equation}
here $\alpha~(\alpha')=1$  or $k$. The eigenvalues problem for the Hamiltonian $H$ is formulated as follows
\begin{equation}\label{h}
H\mid\psi_{1}>=\tilde{E_{1}}\mid\psi_{1}>,
~H\mid\psi_{k}>=E_{k}\mid\psi_{k}>.
\end{equation}
For the eigenstate $\mid\psi_{1}>$ (for small $g$) perturbation
method gives the expression
\begin{equation}\label{f}
\mid\psi_{1}>\approx\mid 1>-\sum_{k}\frac{g
V_{k}}{E_{k}-E_{1}}\mid k>.
\end{equation}
If $E_{1}>0$, Poincare's divergences appear at $E_{k}=E_{1}$.\\ In
accordance with Brussels - Austin group approach the eigenvalues
problem can be solved if the time ordering of the eigenstates will
be introduced. This procedure can be realized through the
introduction into the denominators imaginary terms:
$-i\varepsilon$ for the relaxation processes, which are oriented
into the future and $+i\varepsilon$ for the excitation processes,
which are oriented into the past. In this case the eigenvalues
problem~\eqref{h} is reduced to the complex eigenvalues problem
\begin{align}\label{ht1}
H
\mid\varphi_{1}>=Z_{1}\mid\varphi_{1}>,~<\widetilde{\varphi}_{1}\mid
H = <\widetilde{\varphi}_{1}\mid Z_{1},
\end{align}
\begin{align}\label{ht2}
H\mid\varphi_{k}>=E_{k}\mid\varphi_{k}>,~<\widetilde{\varphi}_{k}\mid
H = <\widetilde{\varphi}_{k}\mid E_{k},
\end{align}
where we must distinguish right - eigenstates $\mid\varphi_{1}>$,
$\mid\varphi_{k}>$ and left - eigenstates
$<\widetilde{\varphi}_{1}\mid$,
$<\widetilde{\varphi}_{k}\mid$~\cite{ps,ppt,opp}, $Z_{1}$ is a
complex: $Z_{1}={\bar{E}_{1}} - i\gamma$, $\bar{E}_{1}$ is a
renormalized energy and $\gamma$ is a real positive value. This
procedure makes it possible to avoid Poincare's divergences and
leads to the following expressions for the eigenstates
$\mid\varphi_{1}>$, $<\widetilde{\varphi}_{1}\mid$~\cite{ppt}
\begin{equation}\label{f43}
\mid\varphi_{1}>\approx\mid 1>-\sum_{k}\frac{g
V_{k}}{(E_{k}-\bar{E}_{1}- z)^{+}_{-i\gamma}}\mid k>,
\end{equation}
\begin{equation}\label{f42}
<\widetilde{\varphi}_{1}\mid\approx<1\mid-\sum_{k}\frac{g
V_{k}}{(E_{k}-\bar{E}_{1}- z)^{+}_{-i\gamma}}<k\mid,
\end{equation}
 In the
expressions~\eqref{f43},~\eqref{f42} the designation
$1/(E_{k}-\bar{E}_{1}- z)^{+}_{-i\gamma}$ has been referred to as
"delayed analytic continuation"~\cite{ppt,opp} and is defined
through the integration with a test function $f(E_{k})$
\begin{equation}\label{ht300}
\begin{split}
\int\limits_{0}^{\infty}dE_{k}\frac{f(E_{k})} {(E_{k}-\bar{E}_{1}-
z)^{+}_{-i\gamma}}\equiv \lim\limits_{z\rightarrow-i\gamma}
\Bigr{(}\int\limits_{0}^{\infty}dE_{k}\frac{f(E_{k})} {E_{k}-
\bar{E}_{1}~-~z}\Bigl{)}_{z\in C^{+}},
\end{split}
\end{equation}
where we first have to evaluate the integration on the upper
half-plane $C^{+}$ and then the limit of $z\rightarrow-i\gamma$
must be taken. \\Now the spectral representation of the
Hamiltonian has the form
\begin{equation}\label{hqw}
H=\sum_{\alpha}Z_{\alpha}\mid\varphi_{\alpha}><{\tilde{\varphi}_{\alpha}}\mid.
\end{equation}
It was shown in ref.~\cite{ppt} that the Hamiltonian~\eqref{hqw} is Hermitian operator. Since $H$ is Hermitian the
corresponding eigenstates $\mid\varphi_{1}>$, $<\widetilde{\varphi}_{1}\mid$ are outside Hilbert space and have no
Hilbert norm
\begin{align}\label{htrr}
&<\varphi_{1}\mid\varphi_{1}>=<{{\tilde{\varphi}}_{1}}\mid{\tilde{\varphi}_{1}}>=0.
\end{align}
For the eigenstates we have relations
\begin{equation}\label{ht}
\sum_{\alpha}\mid\varphi_{\alpha}><{\tilde{\varphi}_{\alpha}}\mid
= 1,~
<{\tilde{\varphi}_{\alpha}}\mid\varphi_{\alpha'}>=\delta_{\alpha\alpha'}.
\end{equation}
The eigenstates $\mid\varphi_{1}>$, $<\widetilde{\varphi}_{1}\mid $ are called "Gamow
vectors"~\cite{gam1}~-~\cite{gam4}.\\ Thus, the Hermiticity of $H$ leads to the fact that "usual" norms of eigenstates
$\mid\varphi_{1}>$, $<\tilde{\varphi}_{1}\mid$ disappear. However, the eigenstates $\mid\varphi_{1}>$,
$<{\tilde{\varphi}_{1}}\mid$ have a broken time symmetry. We can associate $\mid\varphi_{1}>$ with the unstable state,
which vanishes for $t\rightarrow +\infty$, $\mid{\tilde{\varphi}_{1}}>$ corresponds to the state, which vanishes for
$t\rightarrow - \infty$
\begin{equation}\label{hq}
\mid\varphi_{1}(t)>=\exp(-i{\bar{E}_{1}}t-\gamma
t)\mid\varphi_{1}(0)>,
\end{equation}
\begin{equation}\label{hq}
\mid\tilde{\varphi}_{1}(t)>=\exp(-i{\bar{E}_{1}}t+\gamma
t)\mid\tilde{\varphi}_{1}(0)>.
\end{equation}

Obviously, at the present moment, it is very interesting and
necessary to continue further development of the Brussels - Austin
group approach in the framework of the more realistic models of
interaction of the relativistic quantum fields. In the article, I
examine $\pi^{\pm}$ - meson decay such as $\pi^{\pm}\rightarrow
l^{\pm} + \nu_{l}~(\tilde{\nu}_{l})$, where $l=e$ or $\mu$. In
section 2 the definition of the weak interaction model is done. In
section 3 I examine the complex eigenvalues problem - the complex
eigenvalue and eigenstates are obtained. In section 4 on the basis
of the complex representation I consider the Liouville formalism.
Time evolution of the density matrix in the framework of
"subdynamics" approach is determined in section 5. Kinetic, time
irreversible nature of the evolution is shown.

\section{Definition of the weak interaction model}

The Hamiltonian of weak interaction $H_{wk}$ has the form
(determination of weak interaction can be found in
works~\cite{Bil}~-~\cite{wb})
\begin{equation}\label{f1}
H_{wk}=\frac{G}{\sqrt{2}}\int
f_{\pi}\overline{\psi}_{l}\gamma_{\alpha}(1+\gamma_{5})\psi_{\nu}\partial_{\alpha}\varphi_{\pi}d\mathbf{x}
+ H.c.
\end{equation}
In expression~\eqref{f1} $G\approx 10^{-5}/m_{p}^{2}$ ($m_{p}$ -
proton mass), $f_{\pi}=0.91m_{\pi}$ ($m_{\pi}$ - $\pi^{\pm}$-meson
mass). $H.c.$ indicates the Hermitian conjugate. For the operator
of leptons field $\overline{\psi}_{l}$ we have the following
decomposition~\cite{Bil}
\begin{equation}\label{f2}
\begin{split}
&\overline{\psi}_{l}={\overline{\psi}_{l}}
^{(+)}+{\overline{\psi}_{l}}^{(-)}, \\
{\overline{\psi}_{l}}^{(+)}=\frac{1}{(2\pi)^{3/2}}&\int
\Bigl{(}\frac{m_{l}}{E_{{\mathbf{p}}_{l}}
}\Bigr{)}^{1/2}{\overline{u}}^{r_{l}}(-p_{l})e^{ip_{l}x}d_{r_{l}}(p_{l})
d\textbf{p}_{l},
\\
{\overline{\psi}_{l}}^{(-)}=\frac{1}{(2\pi)^{3/2}}&\int
\Bigl{(}\frac{m_{l}}{E_{{\mathbf{p}}_{
l}}}\Bigr{)}^{1/2}\overline{u}^{r_{l}}(p_{l})e^{-ip_{l}x}c^{\dag}_{r_{l}}(p_{l})d\textbf{p}_{l}.
\end{split}
\end{equation}
For the operator of neutrino field we have
\begin{equation}\label{f3}
\begin{split}
{{\psi}_{\nu}}=\frac{1}{(2\pi)^{3/2}}&\int
\bigl{(}{{u}}^{r_{\nu}}(p_{\nu})e^{ip_{\nu}x}c_{r_{\nu}}(p_{\nu})+
{{u}}^{r_{\nu}}(-p_{\nu})e^{-ip_{\nu}x}{d^{\dag}_{r_{\nu}}}(p_{\nu})\bigr{)}
d\textbf{p}_{\nu}.\\
\end{split}
\end{equation}
It is assumed that the neutrino's mass is zero. The operator of
meson field is given by
\begin{equation}\label{f4}
\begin{split}
{{\varphi}_{\pi}}=\frac{1}{(2\pi)^{3/2}}&\int\Bigl{(}\frac{1}{2E_{{\mathbf{p}}_{
\pi}}}\Bigr{)}^{1/2}
\bigl{(}e^{-ip_{\pi}x}a^{\dag}(p_{\pi})+e^{ip_{\pi}x}{b}(p_{\pi})\bigl{)}d\textbf{p}_{\pi}.
\end{split}
\end{equation}
In the decompositions~\eqref{f2}~-~\eqref{f4} such value as $c_{r}(p)$ ($a^{\dag}(p)$, $c^{\dag}_{r}(p)$) is the
operator of destruction (creation) of particle, $d^{\dag}_{r}(p)$ (${b}(p)$, $d_{r}(p)$) is the operator of creation
(destruction) of antiparticle, symbol $"\dag"$ indicates the Hermitian conjugate. Spinors $\bar{u}^{r}(p)$,
$\bar{u}^{r}(-p)$ (${u}^{r}(p)$, ${u}^{r}(-p)$) correspond to the states with helicity $r=\pm 1$, $m_{l}$ - lepton
mass, $E_{\mathbf{p}_{\iota}}=(\mathbf{p}_{\iota}^{2}+m_{\iota}^{2})^{1/2}$ ($\iota=l$~or~$\pi$). Note that we write
4~- vectors in the form $A=(\textbf{A}, iA_{0})$. In this case the following equalities are valid
$A^{2}=\textbf{A}^{2}+A_{4}^{2}=\textbf{A}^{2}-A^{2}_{0}$ and $px\equiv p_{\alpha}x_{\alpha}=\textbf{px}-p_{0}x_{0}$.
We use units with $\hbar$, and the speed of light taken to be unity ($\hbar = c = 1$), $\gamma_{\alpha}$, $\gamma_{5}$
~- Hermitian 4$\times$4 matrices ($\gamma_{\alpha}$$\gamma_{\nu}$ + $\gamma_{\nu}$$\gamma_{\alpha}$ =
2$\delta_{\alpha\nu}$, $\gamma_{\alpha}$$\gamma_{5}$ + $\gamma_{5}$$\gamma_{\alpha}$ = 0, $\gamma^{2}_{5}$ = 1),
$\overline{\psi}_{l}$ = $\psi^{\dag}_{l}$$\gamma_{4}$.

\section{Complex eigenvalues problem - perturbative solutions}
Let examine the eigenvalues problem for the Hamiltonian
$H=H_{0}+H_{wk}$, where $H_{wk}$ - weak interaction. We will solve
the problem assuming that the eigenvalue $Z_{\mathbf{p}_{\pi}}$ of
Hamiltonian $H$ is complex (the general formalism of the complex
spectral representation can be found in
works~\cite{ppt,opp,pop3},~\cite{dfg}~-~\cite{ap3}). In accordance
with the approach~\cite{ppt}, in our case, we will distinguish
equation for the right meson-eigenstate
$\mid\varphi_{\mathbf{p}_{\pi}}>$ and for the left meson-eigenstate
$<\widetilde{\varphi}_{\mathbf{p}_{\pi}}\mid$ of Hamiltonian $H$
\begin{align}\label{ht}
H\mid\varphi_{\mathbf{p}_{\pi}}>=Z_{\mathbf{p}_{\pi}}\mid\varphi_{\mathbf{p}_{\pi}}>,
~<\widetilde{\varphi}_{\mathbf{p}_{\pi}}\mid H =
<\widetilde{\varphi}_{\mathbf{p}_{\pi}}\mid Z_{\mathbf{p}_{\pi}}.
\end{align}
We expand the values $\mid\varphi_{\mathbf{p}_{\pi}}>$,
$<\widetilde{\varphi}_{\mathbf{p}_{\pi}}\mid$,
$Z_{\mathbf{p}_{\pi}}$ in the perturbation series
\begin{align}\label{ht3}
\mid\varphi_{\mathbf{p}_{\pi}}>=\sum\limits_{n=0}^{\infty}g^{n}\mid\varphi_{\mathbf{p}_{\pi}}^{(n)}>,~
<\widetilde{\varphi}_{\mathbf{p}_{\pi}}\mid=\sum\limits_{n=0}^{\infty}g^{n}<\widetilde{\varphi}_{\mathbf{p}_{\pi}}^{(n)}\mid,~
Z_{\mathbf{p}_{\pi}}=\sum\limits_{n=0}^{\infty}g^{n}Z_{\mathbf{p}_{\pi}}^{(n)},
\end{align}
where
\begin{align}\label{ht4}
\mid\varphi_{\mathbf{p}_{\pi}}^{(0)}>=\mid\mathbf{p}_{\pi}>,~
<\widetilde{\varphi}_{\mathbf{p}_{\pi}}^{(0)}\mid=<\mathbf{p}_{\pi}\mid,~
Z_{\mathbf{p}_{\pi}}^{(0)}=E_{\mathbf{p}_{\pi}},~g\equiv G.
\end{align}
Certainty, we will speak about the decay $\pi^{-}\rightarrow \mu^{-}
+ \tilde{\nu}_{\mu}$. In our case one - particle vector
$\mid\mathbf{p}_{\pi}>$ corresponds to the bare $\pi^{-}$ - meson
state with momentum $\mathbf{p}_{\pi}$. Two - particles vector
$\mid\mathbf{p}_{\mu},r_{\mu};\mathbf{p}_{\nu},r_{\nu}>$, where
$\mathbf{p}_{\mu}$, $r_{\mu}$ are momentum and helicity of muon and
$\mathbf{p}_{\nu}$, $r_{\nu}$ are momentum and helicity of neutrino
(in our case antineutrino), corresponds to the bare state consisting
of  muon and neutrino. In the model $\mid\mathbf{p}_{\pi}>$,
$\mid\mathbf{p}_{\mu},r_{\mu};\mathbf{p}_{\nu},r_{\nu}>$ are
eigenstates of the free Hamiltonian $H_{0}$. In accordance with
definitions~\eqref{ht3}~\eqref{ht4}, as will be shown in appendix,
we obtain the expressions
\begin{equation}\label{zzz}
Z_{\mathbf{p}_{\pi}}^{(n)}=<\mathbf{p}_{\pi}\mid V\mid
\varphi_{\mathbf{p}_{\pi}}^{(n-1)}>-\sum^{n-1}_{l=1}Z_{\mathbf{p}_{\pi}}^{(l)}<\mathbf{p}_{\pi}\mid
\varphi_{\mathbf{p}_{\pi}}^{(n-l)}>,
\end{equation}
\begin{equation}\label{f54}
\begin{split}
&<\mathbf{p}_{\mu},r_{\mu};\mathbf{p}_{\nu},r_{\nu}\mid
\varphi_{\mathbf{p}_{\pi}}^{(n)}>=\frac{-1}{E_{\mathbf{p}_{\mu}}+
E_{\mathbf{p}_{\nu}}-E_{\mathbf{p}_{\pi}}+i\varepsilon_{\beta\alpha}}\\
&\times\Bigl{(}<\mathbf{p}_{\mu},r_{\mu};\mathbf{p}_{\nu},r_{\nu}\mid
V\mid
\varphi_{\mathbf{p}_{\pi}}^{(n-1)}>-\sum_{l=1}^{n}Z_{\mathbf{p}_{\pi}}^{(l)}
<\mathbf{p}_{\mu},r_{\mu};\mathbf{p}_{\nu},r_{\nu}\mid
\varphi_{\mathbf{p}_{\pi}}^{(n-l)}>\Bigr{)},
\end{split}
\end{equation}
\begin{equation}\label{f64}
\begin{split}
&<\tilde{\varphi}_{\mathbf{p}_{\pi}}^{(n)}\mid\mathbf{p}_{\mu},r_{\mu};\mathbf{p}_{\nu},r_{\nu}>=
\frac{1}{E_{\mathbf{p}_{\pi}}-
E_{\mathbf{p}_{\mu}}-E_{\mathbf{p}_{\nu}}+i\varepsilon_{\alpha\beta}}\\
&\times\Bigl{(}<\tilde{\varphi}_{\mathbf{p}_{\pi}}^{(n-1)}\mid V\mid
\mathbf{p}_{\mu},r_{\mu};\mathbf{p}_{\nu},r_{\nu}>-\sum_{l=1}^{n}Z_{\mathbf{p}_{\pi}}^{(l)}
<\tilde{\varphi}_{\mathbf{p}_{\pi}}^{(n-l)}\mid
\mathbf{p}_{\mu},r_{\mu};\mathbf{p}_{\nu},r_{\nu}>\Bigr{)},
\end{split}
\end{equation}
where the determination $H_{wk}=GV$ is used (it is necessary to note
that in the paper, for simplification of the expressions,
unessential normalizing volume is implied, but it is not written).
In eqs.~\eqref{f54},~\eqref{f64}, in accordance with the
approach~\cite{ppt}, the time ordering was introduced. Here,
$\varepsilon_{\beta\alpha}$ ($\varepsilon_{\alpha\beta}$) is a
infinitesimal, where $\alpha$ corresponds to the $\pi^{-}$ meson
state, $\beta$ corresponds to the decay products - $\mu^{-}$ and
$\tilde{\nu}_{\mu}$.\\ The quantity
$<\mathbf{p}_{\mu},r_{\mu};\mathbf{p}_{\nu},r_{\nu}\mid
\varphi_{\mathbf{p}_{\pi}}^{(n)}>\equiv <\beta\mid
\varphi_{\alpha}^{(n)}>$ corresponds to the $\alpha\rightarrow\beta$
($\pi^{-}\rightarrow \mu^{-} + \tilde{\nu}_{\mu}$) transition,
$<\tilde{\varphi}_{\mathbf{p}_{\pi}}^{(n)}\mid\mathbf{p}_{\mu},r_{\mu};\mathbf{p}_{\nu},r_{\nu}>\equiv
<\tilde{\varphi}_{\alpha}^{(n)}\mid\beta>$ corresponds to the
$\beta\rightarrow\alpha$ transition. Since unstable $ \pi^{-}$ meson
disappears in the future we associate with the transition
$\alpha\rightarrow\beta$ the analytic continuation
$\varepsilon_{\beta\alpha} = - \varepsilon$. With the reverse
$\beta\rightarrow\alpha$ transition we associate the analytic
continuation oriented to the past, i.e., $\varepsilon_{\alpha\beta}
= + \varepsilon$. In other words  we assume that the state
corresponding to muon and neutrino disappears in the past.
Expressions~\eqref{ht3} -~\eqref{f54} result into~\eqref{apdfrtef}
\begin{equation}\label{f73}
\begin{split}
&Z_{\mathbf{p}_{\pi}}=E_{\mathbf{p}_{\pi}}-\frac{G^{2}f_{\pi}^{2}m_{\mu}}{32{\pi}^{3}E_{\mathbf{p}_{\pi}}}
\sum_{r_{\mu}'r_{\nu}'}\int\frac{d\textbf{p}'_{\mu}d\textbf{p}'_{\nu}}{E_{\mathbf{p}'_{\mu}}}~
\delta(\mathbf{p}_{\pi}-\mathbf{p}'_{\mu}-\mathbf{p}'_{\nu})\\
&\times
\frac{p_{\alpha,\pi}\overline{u}^{r_{\nu}'}(-{p}'_{\nu})\gamma_{\alpha}
(1+\gamma_{5})\delta_{\alpha}u^{r_{\mu}'}(p_{\mu}')
p_{\beta,\pi}\overline{u}^{r_{\mu}'}(p_{\mu}')\gamma_{\beta}
(1+\gamma_{5})u^{r_{\nu}'}(-p_{\nu}')}{(E_{\mathbf{p}'_{\mu}}+E_{\mathbf{p}'_{\nu}}-Z)^{+}_
{Z_{\mathbf{p}_{\pi}}}}~,
\end{split}
\end{equation}
where summation over internal indices $\alpha$, $\beta$ is implied;
$\delta_{\alpha}=-1$ if $\alpha=1,2,3$ and $\delta_{\alpha}=1$ if
$\alpha=4$. Taking into account "delayed analytic
continuation"~\eqref{apperet}
\begin{equation}
\frac{1}{(E_{\mathbf{p}'_{\mu}}+E_{\mathbf{p}'_{\nu}}-Z)^{+}_
{Z_{\mathbf{p}_{\pi}}}}\equiv
\frac{1}{E_{\mathbf{p}'_{\mu}}+E_{\mathbf{p}'_{\nu}}-Z-i\varepsilon}\mid_{Z=Z_{\mathbf{p}_{\pi}}},
\end{equation}
using the formal expression $\frac{1}{w\pm i\varepsilon
}\rightarrow\mathbf{P}\frac{1}{w}\mp i\pi\delta(w)$ and being
limited by order $G^{2}$ we present eq.~\eqref{f73} in the form
\begin{equation}\label{dodo}
Z_{\mathbf{p}_{\pi}}=\bar{E}_{\mathbf{p}_{\pi}}-i\gamma_{\mathbf{p}_{\pi}}.
\end{equation}
In the expression~\eqref{dodo}
\begin{equation}\label{f8}
\begin{split}
&\bar{E}_{\mathbf{p}_{\pi}}=
E_{\mathbf{p}_{\pi}}+\frac{G^{2}f_{\pi}^{2}m_{\mu}}{32{\pi}^{3}E_{\mathbf{p}_{\pi}}}~
\mathbf{P}\sum_{r_{\mu}'r_{\nu}'}\int\frac{d\textbf{p}'_{\mu}d\textbf{p}'_{\nu}}{E_{\mathbf{p}'_{\mu}}}~
\delta(\mathbf{p}_{\pi}-\mathbf{p}'_{\mu}-\mathbf{p}'_{\nu})\\
&\times
\frac{p_{\alpha,\pi}\overline{u}^{r_{\nu}'}(-{p}'_{\nu})\gamma_{\alpha}
(1+\gamma_{5})(-\delta_{\alpha})u^{r_{\mu}'}(p_{\mu}')
p_{\beta,\pi}\overline{u}^{r_{\mu}'}(p_{\mu}')\gamma_{\beta}
(1+\gamma_{5})u^{r_{\nu}'}(-p_{\nu}')}{E_{\mathbf{p}_{\mu}'}+
E_{\mathbf{p}_{\nu}'}-E_{\mathbf{p}_{\pi}}}
\end{split}
\end{equation}
is a renormalized energy of $\pi^{-}$ - meson, $\mathbf{P}$ stands
for the principal part and
\begin{equation}\label{f999}
\begin{split}
&\gamma_{\mathbf{p}_{\pi}}=\frac{G^{2}f_{\pi}^{2}m_{\mu}}{32{\pi}^{2}E_{\mathbf{p}_{\pi}}}~
\sum_{r_{\mu}'r_{\nu}'}\int\frac{d\textbf{p}'_{\mu}d\textbf{p}'_{\nu}}{E_{\mathbf{p}'_{\mu}}}
\delta(\mathbf{p}_{\pi}-\mathbf{p}'_{\mu}-\mathbf{p}'_{\nu})
\delta(E_{\mathbf{p}_{\pi}}-E_{\mathbf{p}'_{\mu}}-E_{\mathbf{p}'_{\nu}})\\
&\times
p_{\alpha,\pi}\overline{u}^{r_{\nu}'}(-{p}'_{\nu})\gamma_{\alpha}
(1+\gamma_{5})\delta_{\alpha}u^{r_{\mu}'}(p_{\mu}')
p_{\beta,\pi}\overline{u}^{r_{\mu}'}(p_{\mu}')\gamma_{\beta}
(1+\gamma_{5})u^{r_{\nu}'}(-p_{\nu}').
\end{split}
\end{equation}
As will be shown in appendix for the  eigenstates $\mid
\varphi_{\mathbf{p}_{\pi}}>$,
$<\tilde{\varphi}_{\mathbf{p}_{\pi}}\mid$ we have
\begin{equation}\label{f10}
\begin{split}
&\mid \varphi_{\mathbf{p}_{\pi}}> = \mid \mathbf{p}_{\pi}> -
i\frac{Gf_{\pi}}{2(2\pi)^{3/2}}\Big{(}\frac{m_{\mu}}{E_{\mathbf{p}_{\pi}}}\Big{)}^{1/2}\\
&\times\sum_{r_{\mu}r_{\nu}}\int\frac{d\mathbf{p}_{\mu}d\mathbf{p}_{\nu}}
{(E_{\mathbf{p}_{\mu}})^{1/2}}~\delta(\mathbf{p}_{\pi}-\mathbf{p}_{\mu}-\mathbf{p}_{\nu})~
\\
&\times\frac{p_{\alpha,\pi}\overline{u}^{r_{\mu}}(p_{\mu})\gamma_{\alpha}
(1+\gamma_{5})u^{r_{\nu}}(-p_{\nu})}
{(E_{\mathbf{p}_{\mu}}+E_{\mathbf{p}_{\nu}}-\bar{E}_{\mathbf{p}_{\pi}}-z)^{+}_
{-i\gamma_{\mathbf{p}_{\pi}}}}
\mid\mathbf{p}_{\mu},r_{\mu};\mathbf{p}_{\nu},r_{\nu}>,
\end{split}
\end{equation}
\begin{equation}\label{f100}
\begin{split}
&<\tilde{\varphi}_{\mathbf{p}_{\pi}}\mid = < \mathbf{p}_{\pi}\mid +
i\frac{Gf_{\pi}}{2(2\pi)^{3/2}}\Big{(}\frac{m_{\mu}}{E_{\mathbf{p}_{\pi}}}\Big{)}^{1/2}\\
&\times\sum_{r_{\mu}r_{\nu}}\int\frac{d\mathbf{p}_{\mu}d\mathbf{p}_{\nu}}{(E_{\mathbf{p}_{\mu}})^{1/2}}~
\delta(\mathbf{p}_{\pi}-\mathbf{p}_{\mu}-\mathbf{p}_{\nu})
\\
&\times\frac{p_{\alpha,\pi}\overline{u}^{r_{\nu}}(-p_{\nu})\gamma_{\alpha}
(1+\gamma_{5})\delta_{\alpha}u^{r_{\mu}}(p_{\mu})}
{(E_{\mathbf{p}_{\mu}}+E_{\mathbf{p}_{\nu}}-\bar{E}_{\mathbf{p}_{\pi}}-z)^{+}_{-i\gamma_{\mathbf{p}_{\pi}}}}
<\mathbf{p}_{\mu},r_{\mu};\mathbf{p}_{\nu},r_{\nu}\mid.
\end{split}
\end{equation}
The eigenstates $\mid\varphi_{\mathbf{p}_{\pi}}>$,
$<{\tilde{\varphi}_{\mathbf{p}_{\pi}}}\mid$ have a broken time
symmetry. In accordance with the
expressions~\eqref{ht},~\eqref{dodo}
\begin{equation}
\mid \varphi_{\mathbf{p}_{\pi}}(t)> = \exp(-iHt)\mid
\varphi_{\mathbf{p}_{\pi}}> = \exp(-i\bar{E}_{\mathbf{p}_{\pi}}t -
\gamma_{\mathbf{p}_{\pi}}t)\mid \varphi_{\mathbf{p}_{\pi}}>.
\end{equation}
\begin{equation}
\mid \tilde{\varphi}_{\mathbf{p}_{\pi}}(t)> = \exp(-iHt)\mid
\tilde{\varphi}_{\mathbf{p}_{\pi}}> =
\exp(-i\bar{E}_{\mathbf{p}_{\pi}}t + \gamma_{\mathbf{p}_{\pi}}t)\mid
\tilde{\varphi}_{\mathbf{p}_{\pi}}>.
\end{equation}
Here $\mid\varphi_{\mathbf{p}_{\pi}}>$ corresponds to the state
which vanishes for $t \rightarrow + \infty$, $\mid
\tilde{\varphi}_{\mathbf{p}_{\pi}}>$ corresponds to the state which
vanishes for $t \rightarrow - \infty$. The states
$\mid\varphi_{\mathbf{p}_{\pi}}>$,
$<{\tilde{\varphi}_{\mathbf{p}_{\pi}}}\mid$ are "Gamow vectors"
(according to classification of~\cite{opp}).
 \\ We
examine $\pi^{-}$ - meson at rest. In this case the
expression~\eqref{f999} results into
\begin{equation}\label{fu100}
\gamma_{\mathbf{p}_{\pi}=0}=\frac{1}{2}\frac{G^{2}f_{\pi}^{2}}{8\pi}m_{\pi}m_{\mu}^{2}\Bigl{(}1-\frac{m_{\mu}^{2}}
{m_{\pi}^{2}}\Bigr{)}^{2}=\frac{1}{2}\Gamma.
\end{equation}
$\Gamma$ in the expression~\eqref{fu100} is well known rate for
$\pi^{-}$ - meson decay (see for example~\cite{wb}). Thus, the
procedure of the time ordering of the expression~\eqref{f54} leads
to the complex eigenvalue of the Hamiltonian $H$ which, in turn,
makes it possible to determine the rate $\Gamma$, when
$1/\Gamma=1/2\gamma_{\mathbf{p}_{\pi}=0}\approx 2.6\times 10^{-8}~
s$ - lifetime $\tau_{0}$ of $\pi^{-}$ - meson at rest.\\
In the general case the value $\gamma_{\mathbf{p}_{\pi}}$ depends
on the momentum $\mathbf{p}_{\pi}$. We examine the situation, when
the angle $\vartheta_{\mathbf{p}_{\pi}}$ of the vector
$\mathbf{p}_{\pi}$ (in the spherical coordinates) is zero. In this
case from the expression~\eqref{f999} we obtain
\begin{equation}\label{fin}
\begin{split}
&\gamma_{|\mathbf{p}|_{\pi}}=\frac{G^{2}f_{\pi}^{2}m^{2}_{\mu}(m^{2}_{\pi}-m^{2}_{\mu})}{16\pi
E_{|\mathbf{p}|_{\pi}}}\\
&\times\int\limits^{1}_{-1}dx\frac{E'_{\mathbf{p}_{\nu}}}{(|\mathbf
{p}|^{2}_{\pi} + m^{2}_{\mu}+ E'^{2}_{\mathbf{p}_{\nu}} -
2|\mathbf {p}|_{\pi}E'_{\mathbf{p}_{\nu}}x )^{1/2} +
E'_{\mathbf{p}_{\nu}} - |\mathbf{p}|_{\pi}x}~,
\end{split}
\end{equation}
where the energy of  neutrino depends on the momentum of $\pi^{-}$
- meson and is determined by the expression
\begin{equation}\label{en}
E'_{\mathbf{p}_{\nu}}=\frac{m^{2}_{\pi}-m^{2}_{\mu}}
{2(E_{|\mathbf{p}|_{\pi}}-|\mathbf{p}|_{\pi}x)}~,~x\equiv
\cos\vartheta_{\mathbf{p}_{\nu}}.
\end{equation}
Note that the obtained results~\eqref{fu100}~-~\eqref{en} are valid
also for $\pi^{+}$ - meson.  As the test of the
expression~\eqref{fin} let me examine the lifetime $\tau$ of
$\pi^{\pm}$ - meson depending on the momentum $|\mathbf{p}|_{\pi}$.
The use of the expression~\eqref{fin} leads to the following
approximate results for the lifetime
$\tau=1/2\gamma_{|\mathbf{p}|_{\pi}}$: $|\mathbf{p}|_{\pi}$ =
0.5~GeV, $\tau \approx 9.8\times 10^{-8}$~s; $|\mathbf{p}|_{\pi}$ =
1.5~GeV, $\tau \approx 2.8\times 10^{-7}$~s; $|\mathbf{p}|_{\pi}$ =
3~GeV, $\tau \approx 5.7\times 10^{-7}$~s~-~the lifetime of
$\pi^{\pm}$ - meson increases with an increasing of the momentum
$|\mathbf{p}|_{\pi}$. Thus, Brussels - Austin group approach leads
to results which are in agreement with Einstein time dilation.\\
In such a way, on the basis of the approach the value of the rate
$\Gamma$ is obtained as the solution of the eigenvalues problem on
the basis of the complex spectral representation.

\section{Liouville formalism in the framework of the complex representation}

In the Liouville formalism  the time evolution is determined by
Liouville - von Neumann equation for the density matrix $\rho$
\begin{equation}\label{ter}
i\frac{\partial\rho(t)}{\partial t}= L\rho(t),
\end{equation}
where "Liouvillian" $L$ has the form
\begin{align}\label{liuvil}
L=H\times 1 - 1\times H,
\end{align}
here symbol "$\times$" denotes the operation ($A\times B$)$\rho=A\rho B$ (the Liouville formalism, for example, can be
found in ref.~\cite{blum}). In accordance with the determination \eqref{liuvil}, $L$ can be written down in the sum of
free part $L_{0}$ that depends on the free Hamiltonian $H_{0}$ and interaction part $L_{I}$ that depends on the
interaction $gV$: $L=L_{0} + L_{I}$. For the Liouville operator $L$ we have the equation (the text is written close to
the materials of works~\cite{ps,opp})
\begin{equation}\label{lv}
L\mid f_{\nu}\rangle\rangle=w_{\nu}\mid f_{\nu}\rangle\rangle,
\end{equation}
where  $\mid
f_{\nu}\rangle\rangle\equiv\mid\psi_{\alpha}><\psi_{\beta}\mid $,
$w_{\nu}=\tilde{E}_{\alpha}-\tilde{E}_{\beta}$ and here $\nu$ is the
correlation index ($\nu$ determines the variety of combinations of
the initial and the final states of the system): $\nu = 0$ if
$\alpha=\beta$ - diagonal case and $\nu\neq 0$ in the remaining
off-diagonal case (the details of the theory of correlations can be
found, for example, in works~\cite{ps,pp3}). The eigenvalues
problem~\eqref{lv} for Liouville operator $L$ has the similar
features as for Hamiltonian $H$. If we expand the values $\mid
f_{\nu}\rangle\rangle$, $w_{\nu}$ in the perturbation series, the
problem of Poincare's divergences will arise again. In accordance
with the approach, we have to introduce the time ordering. This
leads to new formulation of the eigenvalues problem for the operator
$L$
\begin{align}\label{ff3u}
L\mid\Psi^{\nu}_{j}\rangle\rangle=Z^{\nu}_{j}\mid\Psi^{\nu}_{j}\rangle\rangle,~ \langle\langle
\tilde{\Psi}^{\nu}_{j}\mid L=\langle\langle \tilde{\Psi}^{\nu}_{j}\mid Z^{\nu}_{j},
\end{align}
where $Z^{\nu}_{j}$ are the complex values and $j$ is a degeneracy index since one type of correlation index can
correspond to the different states (the complex eigenvalues problem for the Liouville operator is examined in
works~\cite{opp},~\cite{pp3}~-~\cite{pp10}). In eq.~\eqref{ff3u} $L$ is Hermitian. It is possible in case corresponding
eigenstates have no Hilbert norm. For the eigenstates $\mid\Psi^{\nu}_{j}\rangle\rangle$, $\langle\langle
\tilde{\Psi}^{\nu}_{j}\mid$ we have the following biorthogonality and bicompleteness
\begin{align}\label{ff3}
\langle\langle \tilde{\Psi}^{\nu}_{j}\mid\Psi^{\nu'}_{i}\rangle\rangle=\delta_{\nu\nu'}\delta_{ji},~ \sum_{\nu
j}\mid\Psi^{\nu}_{j}\rangle\rangle\langle\langle \tilde{\Psi}^{\nu}_{j}\mid =1.
\end{align}
The spectral representation of the Liouville operator can be
written as follows
\begin{align}\label{ff3}
L=\sum_{\nu j}Z^{\nu}_{j}\mid\Psi^{\nu}_{j}\rangle\rangle\langle\langle \tilde{\Psi}^{\nu}_{j}\mid.
\end{align}
It was shown that the eigenstates of $L$ can be written in the
terms of kinetic operators $C^{\nu}$ and $D^{\nu}$. Operator
$C^{\nu}$ creates correlations other than the $\nu$ correlations,
$D^{\nu}$ is destruction operator~\cite{opp,crit,crit2}. The use
of the kinetic operators allows to write down expressions for the
eigenstates of Liouville operator in the following form~\cite{opp}
\begin{align}\label{wer}
|\Psi^{\nu}_{j}\rangle\rangle=(N^{\nu}_{j})^{1/2}(P^{\nu} +
C^{\nu})|u^{\nu}_{j}\rangle\rangle
,~\langle\langle\widetilde{\Psi}^{\nu}_{j}|=\langle\langle
\widetilde{v}^{\nu}_{j}|(P^{\nu} + D^{\nu})(N^{\nu}_{j})^{1/2},
\end{align}
where $N^{\nu}_{j}$ - is a normalization constant. The
determination of the states $|u^{\nu}_{j}\rangle\rangle$,
$\langle\langle\widetilde{v}^{\nu}_{j}|$ and operators $P^{\nu}$,
$C^{\nu}$, $D^{\nu}$ can be found in works~\cite{opp,oppt,pp3}. In
the general case, for example, the operators $P^{\nu}$ are
determined by the following conditions~\cite{pp3}
\begin{align}\label{ae8}
P^{\nu}=\sum\limits_{j}|u^{\nu}_{j}\rangle\rangle \langle\langle\widetilde{u}^{\nu}_{j}|,
~\langle\langle\widetilde{u}^{\nu}_{j}|u^{\nu'}_{j'}\rangle\rangle=\delta_{\nu\nu'}\delta_{j j'}.
\end{align}
Similarly for $P^{\nu}$ and
$\langle\langle\widetilde{v}^{\nu}_{j}|$ we have:
\begin{align}
P^{\nu}=\sum\limits_{j}|v^{\nu}_{j}\rangle\rangle \langle\langle\widetilde{v}^{\nu}_{j}|,
~\langle\langle\widetilde{v}^{\nu}_{j}|v^{\nu'}_{j'}\rangle\rangle=\delta_{\nu\nu'}\delta_{j j'}.
\end{align}
Substituting expression~\eqref{wer} in eq.~\eqref{ff3u} and multiplying $P^{\nu}$ from left on both sides, we
obtain~\cite{opp}
\begin{align}\label{ae8}
\theta^{\nu}_{C}|u^{\nu}_{j}\rangle\rangle =
Z^{\nu}_{j}|u^{\nu}_{j}\rangle\rangle,
\end{align}
where
\begin{align}\label{ae9}
\theta^{\nu}_{C}\equiv P^{\nu}L(P^{\nu} + C^{\nu})=L_{0}P^{\nu} +
P^{\nu}L_{I}(P^{\nu} + C^{\nu})P^{\nu}.
\end{align}
In eq.~\eqref{ae8} $\theta^{\nu}_{C}$ is the collision operator connected with the kinetic operator $C^{\nu}$. This is
non-Hermitian dissipative operator, which plays the main role in the nonequilibrium dynamics. As was shown in
ref.~\cite{pp3} operator $\theta^{0}_{C}$ can be reduced to the collision operator in Pauli master equation for the
weakly coupled systems. Comparing eqs.~\eqref{ff3u},~\eqref{ae8} we can see that $|u^{\nu}_{j}\rangle\rangle$ is
eigenstate of collision operators $\theta^{\nu}_{C}$ with the same eigenvalues $Z^{\nu}_{j}$ as $L$. It is possible to
obtain the equation for the operator $\theta^{\nu}_{D}$ analogous to eq.~\eqref{ae8}, which is associated
with the destruction kinetic operator $D^{\nu}$.\\
In that way, the determination of the eigenvalues problem for the
Liouville operator $L$ outside the Hilbert space leads to the
connection of quantum mechanics with kinetic, time irreversible
dynamics.

\section{Time evolution of the density matrix}
The fundamental quantum-mechanical Liouville - von Neumann
equation~\eqref{ter} describes the reversible evolution in the time.
Basic question is - how can the irreversibility arise? The response
to the question found its embodiment in the works of Brussels-Austin
group. It was shown that the description of the time irreversible
evolution is possible in the space of $\Psi^{\nu}_{j}$,
$\widetilde{\Psi}^{\nu}_{j}$ - functions. Let introduce the
"subdynamics" approach. The "subdynamics"
approach~\cite{ph}~-~\cite{hgm7} means the construction of the
complete set of the spectral projectors $\Pi^{\nu}$
\begin{align}\label{poper}
\Pi^{\nu}=\sum\limits_{j}|\Psi^{\nu}_{j}\rangle\rangle\langle\langle\widetilde{\Psi}^{\nu}_{j}|.
\end{align}
The projectors $\Pi^{\nu}$ satisfy the following relations:
\begin{equation}\label{com}
\begin{split}
\Pi^{\nu}L=L\Pi^{\nu},~&(\text{commutativity});~
\sum\limits_{\nu}\Pi^{\nu}=1,~(\text{completeness});\\
&\Pi^{\nu}\Pi^{\nu'}=\Pi^{\nu}\delta_{\nu\nu'},~(\text{orthogonality}).
\end{split}
\end{equation}
Thus, the following decomposition of the density matrix is possible
\begin{equation}\label{plo}
\rho(t)=\sum_{\nu}\Pi^{\nu}\rho(t)\equiv\sum_{\nu}\rho^{\nu}(t),
\end{equation}
where $\rho^{\nu}(t)\equiv\Pi^{\nu}\rho(t)$. In the framework of "subdynamics" approach we can reduce eq.~\eqref{ter}
to the equation for $P^{\nu}\rho^{\nu}(t)$ - component for each $\Pi^{\nu}$ - subspace~\cite{opp}
\begin{align}\label{ffiu}
i\frac{\partial}{\partial
t}P^{\nu}\rho^{\nu}=\theta^{\nu}_{C}P^{\nu}\rho^{\nu},
\end{align}
where operator $\theta^{\nu}_{C}$ is determined by the expression~\eqref{ae9}. $P^{\nu}\rho^{\nu}(t)$ - components were
called as the "privileged" components of $\rho^{\nu}(t)$. Projectors $\Pi^{\nu}$ can be associated with the
introduction of the concept of "subdynamics" because the components $\rho^{\nu}$ satisfy separate equations of motion.
Our great interest is to investigate eq.~\eqref{ffiu} for the system of the weak interacting fields. For this purpose
we present the latter in the Dirac - representation
\begin{align}\label{Diracequation}
i\frac{\partial}{\partial
t}P^{\nu}\rho^{\nu}(t)=\vartheta^{\nu}(t)P^{\nu}\rho^{\nu}(t),
\end{align}
where
\begin{align}\label{tet}
\vartheta^{\nu}(t)\equiv P^{\nu}L_{I}(t)C^{\nu}P^{\nu}.
\end{align}
Note that in eq.~\eqref{Diracequation}, the previous designations of operators are preserved. The general solution of
eq.~\eqref{Diracequation} can be found  after examining the equivalent integral equation
\begin{align}\label{Dt}
P^{\nu}\rho^{\nu}(t)=P^{\nu}\rho^{\nu}(t_{0})~
+~(-i)\int\limits^{t}_{t_{0}}dt_{1}\vartheta^{\nu}(t_{1})P^{\nu}\rho^{\nu}(t_{1}),
\end{align}
where $P^{\nu}\rho^{\nu}(t_{0})$ corresponds to the initial moment $t_{0}$. The result can be put down in the form
\begin{align}\label{Dts4}
P^{\nu}\rho^{\nu}(t)=\Omega^{\nu}(t,~t_{0})P^{\nu}\rho^{\nu}(t_{0}),
\end{align}
where
\begin{align}\label{Dts5}
\Omega^{\nu}(t,~t_{0})=\sum\limits^{\infty}_{n=0}(-i)^{n}\int\limits^{t}_{t_{0}}\int\limits^{t_{1}}_{t_{0}}~...~\int\limits^{t_{n-1}}_{t_{0}}
dt_{1}dt_{2}~...~dt_{n}\vartheta^{\nu}(t_{1})\vartheta^{\nu}(t_{2})~...~\vartheta^{\nu}(t_{n}).
\end{align}
The non-Hermitian operator $\Omega^{\nu}(t,~t_{0})$ determines the
time evolution of the "privileged" component $P^{\nu}\rho^{\nu}$ -
the time irreversible evolution of the unstable state. Determination
of eq.~\eqref{Dts4} we will carry out for the matrix element of the
form $\langle\langle
\mathbf{p}_{\pi}~\mathbf{p}_{\pi}|P^{0}\rho^{0}(t)\rangle\rangle\equiv
<\mathbf{p}_{\pi}\mid\rho^{0}(t)\mid\mathbf{p}_{\pi}>\equiv\rho^{0}_{\mathbf{p}_{\pi}
\mathbf{p}_{\pi} }(t)$. Using the determinations of the operators
$P^{\nu}$, $C^{\nu}$, $D^{\nu}$~\cite{opp,oppt,pp3}, for the
diagonal "privileged" component $(\nu = 0)$  (being limited by order
$G^{2}$) from the expression~\eqref{Dts4} we have the result
\begin{equation}\label{fo4}
\begin{split}
&\rho^{0}_{\mathbf{p}_{\pi} \mathbf{p}_{\pi}}(t) =
e^{-2\gamma_{\mathbf{p}_{\pi}} t}\rho^{0}_{\mathbf{p}_{\pi}
\mathbf{p}_{\pi}}(0) + (1-e^{-2\gamma_{\mathbf{p}_{\pi}}
t})\sum_{r_{\mu}r_{\nu}}\int
d\textbf{p}_{\mu}d\textbf{p}_{\nu}\Gamma_{\mathbf{p}_{\pi}
~\mathbf{p}_{\mu},r_{\mu};\mathbf{p}_{\nu},r_{\nu} }\\
&\times\rho^{0}_{~\mathbf{p}_{\mu},r_{\mu};\mathbf{p}_{\nu},r_{\nu}
~\mathbf{p}_{\mu},r_{\mu};\mathbf{p}_{\nu},r_{\nu} }(0),
\end{split}
\end{equation}
where
$t_{0}=0$,~$\rho^{0}_{~\mathbf{p}_{\mu},r_{\mu};\mathbf{p}_{\nu},r_{\nu}
~\mathbf{p}_{\mu},r_{\mu};\mathbf{p}_{\nu},r_{\nu} } \equiv
<\mathbf{p}_{\mu},r_{\mu};\mathbf{p}_{\nu},r_{\nu}\mid\rho^{0}
\mid\mathbf{p}_{\mu},r_{\mu};\mathbf{p}_{\nu},r_{\nu}>$. The result
\eqref{fo4} reflects the evolution from the unstable $\pi^{-}$ -
meson state to the decay products. As follows from
expression~\eqref{fo4} the time evaluation of $\pi^{-}$ - meson has
strictly exponential behavior. Function $\Gamma_{\mathbf{p}_{\pi}
~\mathbf{p}_{\mu},r_{\mu};\mathbf{p}_{\nu},r_{\nu} }$ in
expression~\eqref{fo4} is determined as
\begin{align}\label{Dts900}
\begin{split}
&\Gamma_{\mathbf{p}_{\pi}
~\mathbf{p}_{\mu},r_{\mu};\mathbf{p}_{\nu},r_{\nu} } =
\frac{G^{2}f_{\pi}^{2}m_{\mu}}{32\pi^{3}E_{\mathbf{p}_{\pi}}E_{\mathbf{p}_{\mu}}}~
\delta(\mathbf{p}_{\pi}-\mathbf{p}_{\mu}-\mathbf{p}_{\nu})\\
&\times\frac{p_{\alpha,\pi}\overline{u}^{r_{\nu}}(-{p}_{\nu})\gamma_{\alpha}
(1+\gamma_{5})(-\delta_{\alpha})u^{r_{\mu}}(p_{\mu})
p_{\beta,\pi}\overline{u}^{r_{\mu}}(p_{\mu})\gamma_{\beta}
(1+\gamma_{5})u^{r_{\nu}}(-p_{\nu})}
{(E_{\mathbf{p}_{\mu}}+E_{\mathbf{p}_{\nu}}-\bar{E}_{\mathbf{p}_{\pi}}-z)^{+}_{-i\gamma_{\mathbf{p}_{\pi}}}
(E_{\mathbf{p}_{\mu}}+E_{\mathbf{p}_{\nu}}-\bar{E}_{\mathbf{p}_{\pi}}-z)^{-}_{+i\gamma_{\mathbf{p}_{\pi}}}}~,
\end{split}
\end{align}
where designation
$1/(E_{\mathbf{p}_{\mu}}+E_{\mathbf{p}_{\nu}}-\bar{E}_{\mathbf{p}_{\pi}}-z)^{-}_{+i\gamma_{\mathbf{p}_{\pi}}}
- $ corresponds to the integration, which first of all is carried
out in the lower half complex plane $C^{-}$ and, after that, the
limit of $z\rightarrow +i\gamma_{\mathbf{p}_{\pi}}$ is taken. In
accordance with  works~\cite{opp},~\cite{oppt} we will determine the
function $\Gamma_{\mathbf{p}_{\pi}
~\mathbf{p}_{\mu},r_{\mu};\mathbf{p}_{\nu},r_{\nu} }$ as the line
shape of the $\pi^{-}$ meson decay products. \\There is the direct
connection between eq.~\eqref{Diracequation} and Pauli master
equation. From eq.~\eqref{Diracequation} we obtain
\begin{equation}\label{kin}
\begin{split}
&\frac{\partial\rho^{0}_{\mathbf{p}_{\pi}\mathbf{p}_{\pi}}(t)}{\partial
t}=\sum_{r_{\mu}r_{\nu}}\int
d\mathbf{p}_{\mu}d\mathbf{p}_{\nu}W_{\mathbf{p}_{\pi}~\mathbf{p}_{\mu},r_{\mu};\mathbf{p}_{\nu},r_{\nu}}
\Bigl(\rho^{0}_{\mathbf{p}_{\mu},r_{\mu};\mathbf{p}_{\nu},r_{\nu}~\mathbf{p}_{\mu},r_{\mu};\mathbf{p}_{\nu},r_{\nu}}(t)\\
&-\rho^{0}_{\mathbf{p}_{\pi}\mathbf{p}_{\pi}}(t)\Bigr)
\end{split}
\end{equation}
the analogue of Pauli master equation for $\pi^{-}$ - meson decay,
where the transition rate
$W_{\mathbf{p}_{\pi}~\mathbf{p}_{\mu},r_{\mu};\mathbf{p}_{\nu},r_{\nu}}$
is given by
\begin{equation}
W_{\mathbf{p}_{\pi}~\mathbf{p}_{\mu},r_{\mu};\mathbf{p}_{\nu},r_{\nu}}=
2\gamma_{\mathbf{p}_{\pi}}\Gamma_{\mathbf{p}_{\pi}~\mathbf{p}_{\mu},r_{\mu};\mathbf{p}_{\nu},r_{\nu}}.
\end{equation}
For the function
$W_{\mathbf{p}_{\pi}~\mathbf{p}_{\mu},r_{\mu};\mathbf{p}_{\nu},r_{\nu}}$
the following expression is correct
\begin{equation}
\sum_{r_{\mu}r_{\nu}}\int
d\mathbf{p}_{\mu}d\mathbf{p}_{\nu}W_{\mathbf{p}_{\pi}~\mathbf{p}_{\mu},r_{\mu};\mathbf{p}_{\nu},r_{\nu}}=2\gamma_{\mathbf{p}_{\pi}}.
\end{equation}
Obviously, results~\eqref{fo4},~\eqref{kin} can be interpreted as
follows: diagonal element of the density matrix gives probability to
reveal $\pi^{-}$ - meson with momentum $\mathbf{p}_{\pi}$ at the
moment of time $t$. This probability decreases due to the evolution
in the time to the products consisting of muon and neutrino.
Results~\eqref{fo4},~\eqref{kin} correspond to the kinetic, time
irreversible evolution of the unstable state in the time, which is
oriented into the future. They reflect the energy transfer during
the evolution from unstable $\pi^{-}$ - meson to the decay products
without appearance of the other spontaneous, unstable states. It is
necessary to note, the analogous to the
expressions~\eqref{fo4},~\eqref{kin} results were obtained in
works~\cite{opp} (for the Friedrichs model) and~\cite{pp3}~(for the
potential scattering).

\section{Concluding remarks}

The meson decay has  been investigated long ago (for example in
work~\cite{gtr}). However it is known that the description of
physical world on the  microscopic level, on the basis of
conventional quantum dynamics is defined by the laws of the nature,
which are deterministic and time reversible. The time in the
conventional method (S - matrix approach~\cite{Bil}~-~\cite{wb} )
does not have the chosen direction and the future and the past are
not distinguished. It is obvious that the facts given before are in
the contradiction to our experience, because the world surrounding
us has obvious irreversible nature. In this world the symmetry in
the time is disrupted and the future, and the past play different
roles. Difference between the conventional description of the nature
and those processes in the nature which we observe creates the
conflict situation.  The alternative formulation of quantum dynamics
found its  realization in the works of Brussels-Austin group that
was headed by I. Prigogine. The basic idea of the I. Prigogine and
co-workers is to develop the precise method for the description of
the nature at the macroscopic and microscopic levels, where the
irreversible processes predominate. It was noted that the exact
solution of this problem is impossible on the basis of the
conventional method, unitary principles. Therefore one should speak
about the alternative formulation of the dynamics, that makes it
possible to include the irreversibility in a natural way. In this
connection the studies of the irreversible processes at the
microscopic level - the microscopic formulation of the
irreversibility represents  the special interest. The authors of the
approach deny the conventional opinion that the irreversibility
appears only at the macroscopic level, while the microscopic level
can be described by the laws, reversed in the time. Thus, new
irreversible dynamics with the disrupted symmetry in the time was
formulated. In the approach of Brussels-Austin group the
irreversibility is presented as the property of material itself and
is not defined by the active role of the observer. This approach
implies the passage from the reversible dynamics to the irreversible
time evolution, where the eigenstates have a broken time symmetry.
In the case of "non-integrable" systems the approach leads to the
asymmetry between the past and the future. For $\pi^{\pm}$ meson
decay, on the basis of Brussels-Austin group approach the value of
the rate $\Gamma$ is obtained as the solution of the eigenvalues
problem on the basis of the complex spectral representation. Whereas
the conventional method is based on the set of the well known,
mnemonic rules. The approach contains the important assembling
element. It leads to the unified formulation of quantum and kinetic
dynamics. {\it Let me emphasize that the main purpose of the work is
to describe $\pi^{\pm}$ - meson decay as the irreversible process}.
Despite the fact that this requires a lot of rather lengthy formulas
than the conventional method, where $\pi^{\pm}$ - meson decay is the
time reversible process, it is possible to say: the approach leads
to the adequate description of the evolution in the time of the
relativistic, unstable state, including irreversibility. This
approach clarifies the passage from reversible dynamics to
irreversible time evolution at the microscopic level (one of the
fundamental problems in physics), as it describes an irreversible
process as a rigorous dynamical process in conservative Hamiltonian
systems when the systems are non-integrable in the sense of
Poincare.

\vspace*{7mm}

{\bf Acknowledgements}

I am grateful to Dr. A. A.~Goy, Dr. O. G.~Tkachev for the helpful
suggestions and Dr. A. V.~Molochkov, Dr. D. V.~Shulga for the
support of this work.

\vspace*{7mm}
\setcounter{equation}{0}
\def\theequation{A.\arabic{equation}}
{\bf Appendix. Derivation of the relations}

In appendix, we will obtain the expressions
~\eqref{f73},~\eqref{f10}. Let me introduce the designation $H_{wk}
= GV$. In accordance with the expressions~\eqref{ht},~\eqref{ht3}
the eigenvalues problem for $\mid\varphi_{\mathbf{p}_{\pi}}>$ can be
rewritten in the form
\begin{equation}\label{app1}
(H_{0} +
GV)\sum\limits_{n=0}^{\infty}G^{n}\mid\varphi_{\mathbf{p}_{\pi}}^{(n)}>
=\sum\limits_{n=0}^{\infty}G^{n}Z_{\mathbf{p}_{\pi}}^{(n)}\sum\limits_{n'=0}^
{\infty}G^{n'}\mid\varphi_{\mathbf{p}_{\pi}}^{(n')}>.
\end{equation}
The multiplication of eq.~\eqref{app1} by one-particle vector
$\mid\mathbf{p}_{\pi}>$ leads to the expression:
\begin{equation}\label{appt}
\begin{split}
&<\mathbf{p}_{\pi} \mid\Bigl{(}
H_{0}\sum\limits_{n=0}^{\infty}G^{n}\mid\varphi_{\mathbf{p}_{\pi}}^{(n)}>
+~
GV\sum\limits_{n=0}^{\infty}G^{n}\mid\varphi_{\mathbf{p}_{\pi}}^{(n)}>\Bigr{)} \\
&= <\mathbf{p}_{\pi} \mid
\sum\limits_{n=0}^{\infty}G^{n}Z_{\mathbf{p}_{\pi}}^{(n)}\sum\limits_{n'=0}^
{\infty}G^{n'}\mid\varphi_{\mathbf{p}_{\pi}}^{(n')}>.
\end{split}
\end{equation}
Since, in the model, $\mid\mathbf{p}_{\pi}>$ is the eigenstate of
the free Hamiltonian $H_{0}$: $H_{0}\mid\mathbf{p}_{\pi}> =
E_{\mathbf{p}_{\pi}}\mid\mathbf{p}_{\pi}>$, expression~\eqref{appt}
results into
\begin{equation}\label{appr}
\begin{split}
&E_{\mathbf{p}_{\pi}}\sum\limits_{n=0}^{\infty}G^{n}<\mathbf{p}_{\pi}\mid\varphi_{\mathbf{p}_{\pi}}^{(n)}>
+ \sum\limits_{n=0}^{\infty}G^{n+1}<\mathbf{p}_{\pi}\mid V
\mid\varphi_{\mathbf{p}_{\pi}}^{(n)}> \\
&= E_{\mathbf{p}_{\pi}} +
E_{\mathbf{p}_{\pi}}\sum\limits_{n=1}^{\infty}G^{n}<\mathbf{p}_{\pi}\mid\varphi_{\mathbf{p}_{\pi}}^{(n)}>
+ \sum\limits_{n=1}^{\infty}G^{n}Z_{\mathbf{p}_{\pi}}^{(n)} \\
&+
\sum\limits_{n=1}^{\infty}G^{n}Z_{\mathbf{p}_{\pi}}^{(n)}\sum\limits_{n'=1}^{\infty}G^{n'}
<\mathbf{p}_{\pi}\mid\varphi_{\mathbf{p}_{\pi}}^{(n')}>.
\end{split}
\end{equation}
We can present eq.~\eqref{appr} in the form
\begin{equation}\label{appd}
\begin{split}
&\sum\limits_{n=1}^{\infty}G^{n}<\mathbf{p}_{\pi}\mid V
\mid\varphi_{\mathbf{p}_{\pi}}^{(n-1)}> -
\sum\limits_{n=1}^{\infty}G^{n}Z_{\mathbf{p}_{\pi}}^{(n)}\sum\limits_{n'=0}^{\infty}G^{n'+1}
<\mathbf{p}_{\pi}\mid\varphi_{\mathbf{p}_{\pi}}^{(n'+1)}> \\
&= \sum\limits_{n=1}^{\infty}G^{n}Z_{\mathbf{p}_{\pi}}^{(n)}.
\end{split}
\end{equation}
From the expression~\eqref{appd} we obtain eq.~\eqref{zzz}.\\
The multiplication of eq.~\eqref{app1} by two - particles vector
$\mid\mathbf{p}_{\mu},r_{\mu};\mathbf{p}_{\nu},r_{\nu}>$ gives
\begin{equation}\label{apph}
\begin{split}
&(E_{\mathbf{p}_{\mu}}+
E_{\mathbf{p}_{\nu}})\sum\limits_{n=1}^{\infty}G^{n}
<\mathbf{p}_{\mu},r_{\mu};\mathbf{p}_{\nu},r_{\nu}\mid\varphi_{\mathbf{p}_{\pi}}^{(n)}>
 \\
&+ \sum\limits_{n=0}^{\infty}G^{n+1}
<\mathbf{p}_{\mu},r_{\mu};\mathbf{p}_{\nu},r_{\nu}\mid V\mid
\varphi_{\mathbf{p}_{\pi}}^{(n)}> = \Bigl{(} E_{\mathbf{p}_{\pi}} +
\sum\limits_{n=1}^{\infty}G^{n}Z_{\mathbf{p}_{\pi}}^{(n)}\Bigr{)}\\
&\times\sum\limits_{n'=1}^{\infty}G^{n'}
<\mathbf{p}_{\mu},r_{\mu};\mathbf{p}_{\nu},r_{\nu}\mid\varphi_{\mathbf{p}_{\pi}}^{(n')}>,
\end{split}
\end{equation}
where the determination
$H_{0}\mid\mathbf{p}_{\mu},r_{\mu};\mathbf{p}_{\nu},r_{\nu}> =
(E_{\mathbf{p}_{\mu}}+
E_{\mathbf{p}_{\nu}})\mid\mathbf{p}_{\mu},r_{\mu};\mathbf{p}_{\nu},r_{\nu}>$
was used. Now we can obtain the expression
\begin{equation}\label{apph}
\begin{split}
&\sum\limits_{n =1}^{\infty}G^{n}
<\mathbf{p}_{\mu},r_{\mu};\mathbf{p}_{\nu},r_{\nu}\mid\varphi_{\mathbf{p}_{\pi}}^{(n)}>\\
&=\frac{-1}{E_{\mathbf{p}_{\mu}}+ E_{\mathbf{p}_{\nu}} -
E_{\mathbf{p}_{\pi}}}\Bigl{(}\sum\limits_{n=1}^{\infty}G^{n}
<\mathbf{p}_{\mu},r_{\mu};\mathbf{p}_{\nu},r_{\nu}\mid V\mid
\varphi_{\mathbf{p}_{\pi}}^{(n-1)}> \\
& - \sum\limits_{n =1}^{\infty}\sum\limits_{n'
=0}^{\infty}G^{n}G^{n'}Z_{\mathbf{p}_{\pi}}^{(n)}<\mathbf{p}_{\mu},r_{\mu};\mathbf{p}
_{\nu},r_{\nu}\mid\varphi_{\mathbf{p}_{\pi}}^{(n')}>\Bigr{)},
\end{split}
\end{equation}
which leads to the equation
\begin{equation}\label{appj}
\begin{split}
&<\mathbf{p}_{\mu},r_{\mu};\mathbf{p}_{\nu},r_{\nu}\mid
\varphi_{\mathbf{p}_{\pi}}^{(n)}>\\
&= \frac{-1}{E_{\mathbf{p}_{\mu}}+
E_{\mathbf{p}_{\nu}}-E_{\mathbf{p}_{\pi}}}
\Bigl{(}<\mathbf{p}_{\mu},r_{\mu};\mathbf{p}_{\nu},r_{\nu}\mid V\mid
\varphi_{\mathbf{p}_{\pi}}^{(n-1)}>\\
&-\sum_{l=1}^{n}Z_{\mathbf{p}_{\pi}}^{(l)}
<\mathbf{p}_{\mu},r_{\mu};\mathbf{p}_{\nu},r_{\nu}\mid
\varphi_{\mathbf{p}_{\pi}}^{(n-l)}>\Bigr{)}.
\end{split}
\end{equation}
In accordance with Brussels - Austin group approach the time
ordering of eq.~\eqref{appj} must be introduced. This can be
realized in accordance with the rules of section {\bf 3} through the
introduction into the denominator imaginary term $-i\varepsilon$
with respect to one of the particle $\mu^{-}$ or $\tilde{\nu}_{\mu}$
(we assume that $\mu^{-}$ and $\tilde{\nu}_{\mu}$ appear in the
future), then we obtain eq.~\eqref{f54}.\\
Now we examine the expression~\eqref{zzz}. We define:
$\mid\mathbf{p}_{\pi}> = b^{\dag}(p_{\pi})\mid\Phi_{0}>$ and
$\mid\mathbf{p}_{\mu},r_{\mu};\mathbf{p}_{\nu},r_{\nu}> =
c^{\dag}_{r_{\mu}}(p_{\mu})d^{\dag}_{r_{\nu}}(p_{\nu})\mid\Phi_{0}>$,
where $b^{\dag}(p_{\pi})$, $c^{\dag}_{r_{\mu}}(p_{\mu})$,
$d^{\dag}_{r_{\nu}}(p_{\nu})$ are the creation operators of
$\pi^{-}$, $\mu^{-}$, $\tilde{\nu}_{\mu}$ particles respectively,
$\mid\Phi_{0}>$ - the vacuum state. Substituting the
expressions~\eqref{f1}~-~\eqref{f4} into the first term of
eq.~\eqref{zzz} (where $V=H_{wk}/G$) we obtain
\begin{equation}\label{appc}
\begin{split}
&Z_{\mathbf{p}_{\pi}}^{(n)} = -i\frac{f_{\pi}}{2{(2\pi)}^{3/2}}
\sum_{r_{\mu}'r_{\nu}'}\int
d\textbf{p}'_{\mu}d\textbf{p}'_{\nu}~\sqrt{\frac{m_{\mu}}
{E_{\mathbf{p}'_{\mu}}E_{\mathbf{p}_{\pi}}}} ~
\delta(\mathbf{p}_{\pi}-\mathbf{p}'_{\mu}-\mathbf{p}'_{\nu})\\
&\times e^{-i(E_{\mathbf{p}'_{\mu}}+
E_{\mathbf{p}'_{\nu}}-E_{\mathbf{p}_{\pi}})t}
p_{\alpha,\pi}\overline{u}^{r_{\nu}'}(-{p}'_{\nu})\gamma_{\alpha}
(1+\gamma_{5})\delta_{\alpha}u^{r_{\mu}'}(p_{\mu}')\\
&\times <\mathbf{p}'_{\mu},r'_{\mu};\mathbf{p}'_{\nu},r'_{\nu}\mid
\varphi_{\mathbf{p}_{\pi}}^{(n-1)}>
-\sum^{n-1}_{l=1}Z_{\mathbf{p}_{\pi}}^{(l)}<\mathbf{p}_{\pi}\mid
\varphi_{\mathbf{p}_{\pi}}^{(n-l)}>,
\end{split}
\end{equation}
where summation over internal index $\alpha$ is implied;
$\delta_{\alpha}=-1$ if $\alpha=1,2,3$ and $\delta_{\alpha}=1$ if
$\alpha=4$. By multiplying eq.~\eqref{appc} by $G^{n}$ and summing
with respect to $n$, we have
\begin{equation}\label{apper}
\begin{split}
&Z_{\mathbf{p}_{\pi}} =
E_{\mathbf{p}_{\pi}}-i\frac{f_{\pi}G}{2{(2\pi)}^{3/2}}
\sum_{r_{\mu}'r_{\nu}'}\int
d\textbf{p}'_{\mu}d\textbf{p}'_{\nu}~\sqrt{\frac{m_{\mu}}
{E_{\mathbf{p}'_{\mu}}E_{\mathbf{p}_{\pi}}}} ~
\delta(\mathbf{p}_{\pi}-\mathbf{p}'_{\mu}-\mathbf{p}'_{\nu})\\
&\times e^{-i(E_{\mathbf{p}'_{\mu}}+
E_{\mathbf{p}'_{\nu}}-E_{\mathbf{p}_{\pi}})t}
p_{\alpha,\pi}\overline{u}^{r_{\nu}'}(-{p}'_{\nu})\gamma_{\alpha}
(1+\gamma_{5})\delta_{\alpha}u^{r_{\mu}'}(p_{\mu}')\\
&\times\frac{<\mathbf{p}'_{\mu},r'_{\mu};\mathbf{p}'_{\nu},r'_{\nu}\mid
\varphi_{\mathbf{p}_{\pi}}>}{<\mathbf{p}_{\pi}
\mid\varphi_{\mathbf{p}_{\pi}}>}.
\end{split}
\end{equation}
Expression for
$<\mathbf{p'}_{\mu},r'_{\mu};\mathbf{p'}_{\nu},r'_{\nu}\mid
\varphi_{\mathbf{p}_{\pi}}>$ can be obtained from eq.~\eqref{f54}.
Using~\eqref{f1}~-~\eqref{f4} we find
\begin{equation}\label{appeu}
\begin{split}
&<\mathbf{p}'_{\mu},r'_{\mu};\mathbf{p}'_{\nu},r'_{\nu}\mid
\varphi_{\mathbf{p}_{\pi}}^{(n)}> = \frac{-1}{E_{\mathbf{p}'_{\mu}}+
E_{\mathbf{p}'_{\nu}}-E_{\mathbf{p}_{\pi}}-i\varepsilon}\\
&\times i\frac{f_{\pi}}{2{(2\pi)}^{3/2}}\Bigl{(}\int
d\textbf{p}'_{\pi}\sqrt{\frac{m_{\mu}}
{E_{\mathbf{p}'_{\mu}}E_{\mathbf{p}'_{\pi}}}} ~
\delta(\mathbf{p}'_{\pi}-\mathbf{p}'_{\mu}-\mathbf{p}'_{\nu})\\
&\times e^{i(E_{\mathbf{p}'_{\mu}}+
E_{\mathbf{p}'_{\nu}}-E_{\mathbf{p}'_{\pi}})t}p~'_{\beta,\pi}\overline{u}^{r_{\mu}'}(p_{\mu}')\gamma_{\beta}
(1+\gamma_{5})u^{r_{\nu}'}(-p_{\nu}')\\
&\times <\mathbf{p}'_{\pi} \mid\varphi_{\mathbf{p}_{\pi}}^{(n-1)}> -
\sum_{l=1}^{n}Z_{\mathbf{p}_{\pi}}^{(l)}
<\mathbf{p}'_{\mu},r'_{\mu};\mathbf{p}'_{\nu},r'_{\nu}\mid
\varphi_{\mathbf{p}_{\pi}}^{(n-l)}>\Bigr{)},
\end{split}
\end{equation}
where summation over internal index $\beta$ is implied. By
multiplying $G^{n}$ to eq.~\eqref{appeu} and summing with respect to
$n$, we get
\begin{equation}\label{apphjh}
\begin{split}
&<\mathbf{p}'_{\mu},r'_{\mu};\mathbf{p}'_{\nu},r'_{\nu}\mid
\varphi_{\mathbf{p}_{\pi}}> = \frac{-1}{E_{\mathbf{p}'_{\mu}}+
E_{\mathbf{p}'_{\nu}}-E_{\mathbf{p}_{\pi}}-i\varepsilon}\\
&\times i\frac{f_{\pi}G}{2{(2\pi)}^{3/2}}\Bigl{(}\int
d\textbf{p}'_{\pi}\sqrt{\frac{m_{\mu}}
{E_{\mathbf{p}'_{\mu}}E_{\mathbf{p}'_{\pi}}}} ~
\delta(\mathbf{p}'_{\pi}-\mathbf{p}'_{\mu}-\mathbf{p}'_{\nu})\\
&\times e^{i(E_{\mathbf{p}'_{\mu}}+
E_{\mathbf{p}'_{\nu}}-E_{\mathbf{p}'_{\pi}})t}p~'_{\beta,\pi}\overline{u}^{r_{\mu}'}(p_{\mu}')\gamma_{\beta}
(1+\gamma_{5})u^{r_{\nu}'}(-p_{\nu}')\\
&\times <\mathbf{p}'_{\pi} \mid\varphi_{\mathbf{p}_{\pi}}> +~
(E_{\mathbf{p}_{\pi}} - Z_{\mathbf{p}_{\pi}})
<\mathbf{p}'_{\mu},r'_{\mu};\mathbf{p}'_{\nu},r'_{\nu}\mid
\varphi_{\mathbf{p}_{\pi}}>\Bigr{)}.
\end{split}
\end{equation}
Expression~\eqref{apphjh} can be represented in the form
\begin{equation}\label{apfgh}
\begin{split}
&<\mathbf{p}'_{\mu},r'_{\mu};\mathbf{p}'_{\nu},r'_{\nu}\mid
\varphi_{\mathbf{p}_{\pi}}> = -
i\frac{f_{\pi}G}{2{(2\pi)}^{3/2}}\int
d\textbf{p}'_{\pi}\sqrt{\frac{m_{\mu}}
{E_{\mathbf{p}'_{\mu}}E_{\mathbf{p}'_{\pi}}}} \\
&\times\delta(\mathbf{p}'_{\pi}-\mathbf{p}'_{\mu}-\mathbf{p}'_{\nu})
e^{i(E_{\mathbf{p}'_{\mu}}+
E_{\mathbf{p}'_{\nu}}-E_{\mathbf{p}'_{\pi}})t}p~'_{\beta,\pi}\overline{u}^{r_{\mu}'}(p_{\mu}')\gamma_{\beta}
(1+\gamma_{5})u^{r_{\nu}'}(-p_{\nu}')\\
&\times <\mathbf{p}'_{\pi}
\mid\varphi_{\mathbf{p}_{\pi}}>\sum_{n=0}^{\infty}
\frac{(Z_{\mathbf{p}_{\pi}} - E_{\mathbf{p} _{\pi}})^{n}}{(
E_{\mathbf{p}'_{\mu}}+
E_{\mathbf{p}'_{\nu}}-E_{\mathbf{p}_{\pi}}-i\varepsilon)^{n+1}}.
\end{split}
\end{equation}
The substitution of result~\eqref{apfgh} into~\eqref{apper} gives
\begin{equation}\label{f73er}
\begin{split}
&Z_{\mathbf{p}_{\pi}}=E_{\mathbf{p}_{\pi}}-\frac{G^{2}f_{\pi}^{2}m_{\mu}}{32{\pi}^{3}E_{\mathbf{p}_{\pi}}}
\sum_{r_{\mu}'r_{\nu}'}\int\frac{d\textbf{p}'_{\mu}d\textbf{p}'_{\nu}}{E_{\mathbf{p}'_{\mu}}}~
\delta(\mathbf{p}_{\pi}-\mathbf{p}'_{\mu}-\mathbf{p}'_{\nu})\\
&\times
p_{\alpha,\pi}\overline{u}^{r_{\nu}'}(-{p}'_{\nu})\gamma_{\alpha}
(1+\gamma_{5})\delta_{\alpha}u^{r_{\mu}'}(p_{\mu}')
p_{\beta,\pi}\overline{u}^{r_{\mu}'}(p_{\mu}')\gamma_{\beta}
(1+\gamma_{5})u^{r_{\nu}'}(-p_{\nu}')\\
&\times \sum_{n=0}^{\infty} \frac{(Z_{\mathbf{p}_{\pi}} -
E_{\mathbf{p} _{\pi}})^{n}}{( E_{\mathbf{p}'_{\mu}}+
E_{\mathbf{p}'_{\nu}}-E_{\mathbf{p}_{\pi}}-i\varepsilon)^{n+1}},
\end{split}
\end{equation}
where the order $G^{2}$  is preserved (for this purpose  the
relationship $<\mathbf{p}'_{\pi} \mid\varphi_{\mathbf{p}_{\pi}}> =
\delta(\mathbf{p}'_{\pi} - \mathbf{p}_{\pi})$ was used). Here, for
the simplification of the intermediate expressions unessential
normalizing volume was not written, but it was implied. Now we
examine the sum
\begin{equation}\label{appmnb}
\begin{split}
\sum_{n=0}^{\infty} \frac{(Z_{\mathbf{p}_{\pi}} - E_{\mathbf{p}
_{\pi}})^{n}}{( E_{\mathbf{p}'_{\mu}}+
E_{\mathbf{p}'_{\nu}}-E_{\mathbf{p}_{\pi}}-i\varepsilon)^{n+1}}.
\end{split}
\end{equation}
If we sum up the series without paying attention to $i\varepsilon$,
we have
\begin{equation}\label{apdfg}
\begin{split}
\sum_{n=0}^{\infty} \frac{(Z_{\mathbf{p}_{\pi}} -
E_{\mathbf{p}_{\pi}})^{n}}{( E_{\mathbf{p}'_{\mu}}+
E_{\mathbf{p}'_{\nu}}-E_{\mathbf{p}_{\pi}})^{n+1}} =
\frac{1}{E_{\mathbf{p}'_{\mu}}+ E_{\mathbf{p}'_{\nu}} -
Z_{\mathbf{p}_{\pi}}}.
\end{split}
\end{equation}
Expression~\eqref{apdfg} has a pole in the lower half plane because,
as it can be shown, $Z_{\mathbf{p}_{\pi}}$ is in the lower half
plane. However each term of the sum eq.~\eqref{appmnb} has a pole at
$E_{\mathbf{p}'_{\mu}} = E_{\mathbf{p}_{\pi}} -
E_{\mathbf{p}'_{\nu}} + i\varepsilon$ or $E_{\mathbf{p}'_{\nu}} =
E_{\mathbf{p}_{\pi}} - E_{\mathbf{p}'_{\mu}} + i\varepsilon$ in the
upper half plane. (the last expressions are equivalent since, both
$\mu^{-}$ and $\tilde{\nu}_{\mu}$ particles are located in the
future with respect to the $\pi^{-}$ meson). This implies that the
summation introduces a discontinuity. To avoid this difficulty we
will adhere to the rule, which was proposed in work~\cite{ppt}
\begin{equation}\label{apperet}
\begin{split}
\sum_{n=0}^{\infty} \frac{(Z_{\mathbf{p}_{\pi}} - E_{\mathbf{p}
_{\pi}})^{n}}{( E_{\mathbf{p}'_{\mu}}+
E_{\mathbf{p}'_{\nu}}-E_{\mathbf{p}_{\pi}}-i\varepsilon)^{n+1}} =
\frac{1} {(E_{\mathbf{p}'_{\mu}}+E_{\mathbf{p}'_{\nu}}-Z)^{+}_
{Z_{\mathbf{p}_{\pi}}}}.
\end{split}
\end{equation}
In work~\cite{ppt} this procedure was named as "delayed analytic
continuation", where we first have to evaluate the integration on
the upper half-plane of $Z$, designated as "+" (with respect to
$\mu^{-}$ or $\tilde{\nu}_{\mu}$ particle), and then substitute
$Z=Z_{\mathbf{p}_{\pi}}$ Using~\eqref{f73er},~\eqref{apperet} we
find eq.~\eqref{f73}
\begin{equation}\label{apdfrtef}
\begin{split}
&Z_{\mathbf{p}_{\pi}}=E_{\mathbf{p}_{\pi}}-\frac{G^{2}f_{\pi}^{2}m_{\mu}}{32{\pi}^{3}E_{\mathbf{p}_{\pi}}}
\sum_{r_{\mu}'r_{\nu}'}\int\frac{d\textbf{p}'_{\mu}d\textbf{p}'_{\nu}}{E_{\mathbf{p}'_{\mu}}}~
\delta(\mathbf{p}_{\pi}-\mathbf{p}'_{\mu}-\mathbf{p}'_{\nu})\\
&\times
\frac{p_{\alpha,\pi}\overline{u}^{r_{\nu}'}(-{p}'_{\nu})\gamma_{\alpha}
(1+\gamma_{5})\delta_{\alpha}u^{r_{\mu}'}(p_{\mu}')
p_{\beta,\pi}\overline{u}^{r_{\mu}'}(p_{\mu}')\gamma_{\beta}
(1+\gamma_{5})u^{r_{\nu}'}(-p_{\nu}')}{(E_{\mathbf{p}'_{\mu}}+E_{\mathbf{p}'_{\nu}}-Z)^{+}_
{Z_{\mathbf{p}_{\pi}}}}.
\end{split}
\end{equation}
Now we obtain the right-eigenstate~\eqref{f10}. We expand
eigenvector $\mid \varphi_{\mathbf{p}_{\pi}}>$ in the terms of the
set of eigenvectors $\mid\mathbf{p}_{\pi}>$,
$\mid\mathbf{p}_{\mu},r_{\mu};\mathbf{p}_{\nu},r_{\nu}>$ of the
Hamiltonian $H_{0}$
\begin{equation}\label{apdfrrty}
\begin{split}
&\mid \varphi_{\mathbf{p}_{\pi}}> =
\int\mid\mathbf{p}'_{\pi}><\mathbf{p}'_{\pi}\mid\varphi_{\mathbf{p}_{\pi}}>d\mathbf{p}'_{\pi}\\
&+ \sum_{r_{\mu}'r_{\nu}'}\int
\mid\mathbf{p}'_{\mu},r'_{\mu};\mathbf{p}'_{\nu},r'_{\nu}><\mathbf{p}'_{\mu},r'_{\mu};\mathbf{p}'_{\nu},r'_{\nu}\mid
\varphi_{\mathbf{p}_{\pi}}>d\textbf{p}'_{\mu}d\textbf{p}'_{\nu}.
\end{split}
\end{equation}
Using the energy conservation law for $\pi^{-}$ meson decay,
$E_{\mathbf{p}_{\pi}} = E_{\mathbf{p}_{\mu}}+E_{\mathbf{p}_{\nu}}$,
with the help of~\eqref{apfgh},~\eqref{apperet} we obtain
\begin{equation}\label{appabc}
\begin{split}
&\mid \varphi_{\mathbf{p}_{\pi}}> =
\int<\mathbf{p}'_{\pi}\mid\varphi_{\mathbf{p}_{\pi}}>\Bigl{(}\mid\mathbf{p}'_{\pi}> - \\
&- \sum_{r_{\mu}'r_{\nu}'}\int
d\textbf{p}'_{\mu}d\textbf{p}'_{\nu}\frac{if_{\pi}G}{2{(2\pi)}^{3/2}}
\sqrt{\frac{m_{\mu}} {E_{\mathbf{p}'_{\mu}}E_{\mathbf{p}'_{\pi}}}}~
\delta(\mathbf{p}'_{\pi}-\mathbf{p}'_{\mu}-\mathbf{p}'_{\nu})\\
&\times\frac{
p~'_{\beta,\pi}\overline{u}^{r_{\mu}'}(p_{\mu}')\gamma_{\beta}
(1+\gamma_{5})u^{r_{\nu}'}(-p_{\nu}')}{
(E_{\mathbf{p}'_{\mu}}+E_{\mathbf{p}'_{\nu}}-Z)^{+}_
{Z_{\mathbf{p}_{\pi}}} }
\mid\mathbf{p}'_{\mu},r'_{\mu};\mathbf{p}'_{\nu},r'_{\nu}>\Bigr{)}d\textbf{p}'_{\pi}.
\end{split}
\end{equation}
The result~\eqref{dodo} makes it possible to rewrite
expression~\eqref{apperet} in the form
\begin{equation}\label{apsdfgh}
\begin{split}
\frac{1} {(E_{\mathbf{p}'_{\mu}}+E_{\mathbf{p}'_{\nu}}-Z)^{+}_
{Z_{\mathbf{p}_{\pi}}}}\equiv \frac{1}
{(E_{\mathbf{p}'_{\mu}}+E_{\mathbf{p}'_{\nu}}-\bar{E}_{\mathbf{p}_{\pi}}-z)^{+}_
{-i\gamma_{\mathbf{p}_{\pi}}}}
\end{split}
\end{equation}
The function~\eqref{apsdfgh} is defined through the integration over
$E_{\mathbf{p}'_{\mu}}$ or $E_{\mathbf{p}'_{\nu}}$
\begin{equation}\label{hddfgh}
\begin{split}
&\int\limits_{0}^{\infty}dE_{\mathbf{p}'_{\mu,(\nu)}}\frac{f(E_{\mathbf{p}'_{\mu,(\nu)}})}
{(E_{\mathbf{p}'_{\mu}}+E_{\mathbf{p}'_{\nu}}-\bar{E}_{\mathbf{p}_{\pi}}-z)^{+}_
{-i\gamma_{\mathbf{p}_{\pi}}}}\\
&\equiv \lim\limits_{z\rightarrow-i\gamma_{\mathbf{p}_{\pi}}}
\Bigr{(}\int\limits_{0}^{\infty}
dE_{\mathbf{p}'_{\nu,(\mu)}}\frac{f(E_{\mathbf{p}'_{\mu,(\nu)}})}
{(E_{\mathbf{p}'_{\mu}}+E_{\mathbf{p}'_{\nu}}-\bar{E}_{\mathbf{p}_{\pi}}-z)_{z\in
C^{+}}}\Bigl{)},
\end{split}
\end{equation}
where we first have to evaluate the integration on the upper
half-plane $C^{+}$ (with respect to $E_{\mathbf{p}'_{\mu}}$ or
$E_{\mathbf{p}'_{\nu}}$) and then the limit of
$z\rightarrow-i\gamma_{\mathbf{p}_{\pi}}$ must be taken,
$f(E_{\mathbf{p}'_{\mu,(\nu)}})$ is a test function. The
expressions~\eqref{appabc}~-~\eqref{hddfgh} give the
result~\eqref{f10}.

\end{document}